\documentstyle[aps,psfig,multicol]{revtex}
\newcommand{\al}{\alpha}
\newcommand{\vE}{{\vec E}}
\newcommand{\r}{{r}}
\newcommand{\rp}{{r^\prime}}
\newcommand{\rdp}{{r''}}
\newcommand{\nr}{n_\r}
\newcommand{\lr}{l_\r}
\newcommand{\e}{\hat{e}}
\newcommand{\enu}{\e_\nu}
\newcommand{\epsnl}{\epsilon(\nr|\lr)}
\newcommand{\ea}{\epsilon_a}
\newcommand{\eb}{\epsilon_b}
\newcommand{\epo}{\epsilon_0}
\newcommand{\ar}{\alpha_\r}
\newcommand{\br}{\beta_\r}

\newcommand{\calC}{{\cal C}}
\newcommand{\cprime}{{\cal{C}^{\prime}}}
\newcommand{\cdprime}{{\calC^{\prime\prime}}}

\newcommand{\ndp}{n^{\prime\prime}}
\newcommand{\ccp}{\calC \rightarrow \cprime}
\newcommand{\cdc}{\cdprime \rightarrow \calC}

\newcommand{\rx}{\rho_x}
\newcommand{\ry}{\rho_y}

\newcommand{\rt}{\tilde\rho}
\newcommand{\rtx}{\partial_x \tilde\rho}

\newcommand{\rtt}{\partial_t \tilde\rho}
\newcommand{\htx}{\partial_x h}
\newcommand{\htxx}{\partial_{xx} h}
\newcommand{\htt}{\partial_t h}

\newcommand{\Rbar}{\overline{R}}
\newcommand{\Cbar}{\overline{\calC}}
\newcommand{\Cijbar}{\overline{\calC^{ij}}}
\newcommand{\Wbar}{\overline{W}}
\newcommand{\half}{{1\over 2}}

\begin{document}

\title{Driven Lattice Gases with Quenched Disorder: Exact Results and
Different Macroscopic Regimes}
\author{Goutam Tripathy and Mustansir Barma}
\address{Theoretical Physics Group, Tata Institute of Fundamental Research,
Homi Bhabha Road, Mumbai 400 005}
\maketitle
\centerline{\today}

\begin{abstract}
 
  We study the effect of quenched spatial disorder on the steady states of
driven systems of interacting particles.  Two sorts of models are studied:
disordered drop-push processes and their generalizations, and the
disordered asymmetric simple exclusion process. We write down the exact
steady-state measure, and consequently a number of physical quantities
explicitly, for the drop-push dynamics in any dimensions for arbitrary
disorder. We find that three qualitatively different regimes of behaviour
are possible in 1-$d$ disordered driven systems. In the Vanishing-Current
regime, the steady-state current approaches zero in the thermodynamic
limit. A system with a non-zero current can either be in the Homogeneous
regime, chracterized by a single macroscopic density, or the
Segregated-Density regime, with macroscopic regions of different
densities. We comment on certain important constraints to be taken care of
in any field theory of disordered systems.
 

\end{abstract}
\begin{multicols}{2}
\section{Introduction}
\label{sec:introduction}
      It is known that quenched disorder can strongly affect the
      large-scale, long-time behaviour of nonequilibrium driven systems
      with interacting constituents.  The interplay of disorder,
      interactions and drive opens up the possibility of new regimes of
      complex and interesting behaviour arising in these systems
      \cite{Fisher}.  In the theoretical effort to delineate and explore
      regimes of new behaviour, an important role is played by simple
      models which capture some features of more complex physical systems.
      In this paper, we study disordered driven diffusive systems by
      analysing stochastically evolving lattice gas models, with quenched
      disordered hopping rates \cite {prl97}. 

      Driven diffusive systems in the absence of disorder have been studied
      extensively and are reviewed in \cite{DDS}.  Also, systems with
      disorder and drive but no interactions between particles are well
      studied and understood \cite{BouGeo}.  But there have been only
      sporadic studies of disordered driven diffusive systems of
      interacting particles.  It has been argued that strong enough random
      site dilution can substantially affect the transport properties of
      particles with hard-core interactions, and can make the system
      respond nonmonotonically to the driving field \cite{RB,BR}.  On the
      other hand, a low concentration of blocked sites was found
      numerically not to affect the critical behaviour of a driven lattice
      gas with additional attractive inter-particle interactions
      \cite{Lauritsen}.  Finally, a driven lattice gas with a quenched
      noise distribution was studied using field-theoretic techniques in
      \cite{Janssen}, but the connection of this study with
      particle-conserving disordered lattice gas models is not clear.

      In this paper, we study disordered lattice gas models with a view
      towards identifying different sorts of generic behaviour that can
      arise on large scales as a consequence of disorder.  The only
      interaction included is the hard-core constraint which limits the
      allowed occupancy of each site.  Our results pertain mostly, but not
      exclusively, to one dimension.  In the remainder of this
      Introduction, we discuss the different types of behaviour displayed
      by the lattice gas models under study.

      We find three distinct regimes in disordered driven diffusive systems
      in one dimension:

      In the {\it Homogeneous} regime, the state of the system is
      characterized by a single density and a nonzero current.  Quenched
      disorder induces variations of the density on the microscopic scale,
      of the order of a few lattice spacings. However, the system has a
      macroscopically homogeneous density. In the thermodynamic
      limit, the current approaches a finite value.

      In the {\it Segregated-Density} regime, the state of the system is
      characterized by two distinct values of density, and a nonzero
      current.  Besides microscopic-scale variations of the density, there
      are  macroscopic regions with differing high and low
      densities. The state is thus characterized by phase separation of
      the density, and a spatially constant time-averaged current which
      remains finite in the thermodynamic limit.

      In the {\it Vanishing-Current} regime, the state of the system is
      characterized by two distinct values of the density, and an
      essentially zero current.  The hallmark of this regime is that the
      current decreases as the system size increases, and vanishes in
      the thermodynamic limit. This is a consequence of rare but
      rate-limiting backbends, or stretches of bonds which disfavour the
      forward flow of current.  The density is inhomogeneous on a
      macroscopic scale.

      The density profiles in typical states in each of the three regimes
      are depicted in Figure 1, while Figure 2 shows the variation of the
      current with system size in the three cases.

\begin{figure}[tb]
\begin{center}
\leavevmode
\psfig{figure=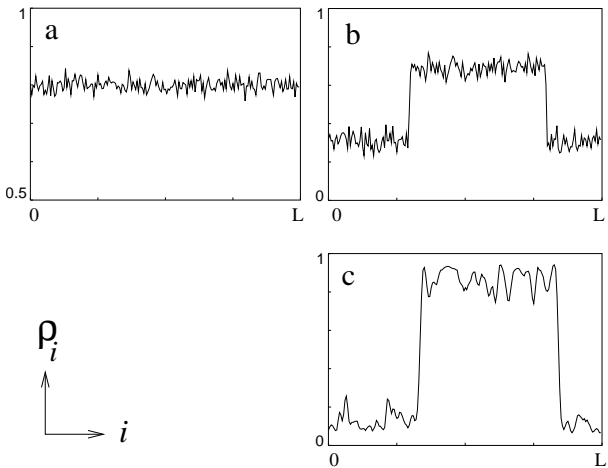,width=8cm}
\end{center}
\narrowtext
\caption{Representative steady-state density profiles 
for the (a) Homogeneous (b) Segregated-Density and (c) Vanishing-Current
regimes in the Disordered Asymmetric Exclusion Process (DASEP). }
\label{fig:regimes1}
\end{figure}
\begin{figure}[tb]
\begin{center}
\leavevmode
\psfig{figure=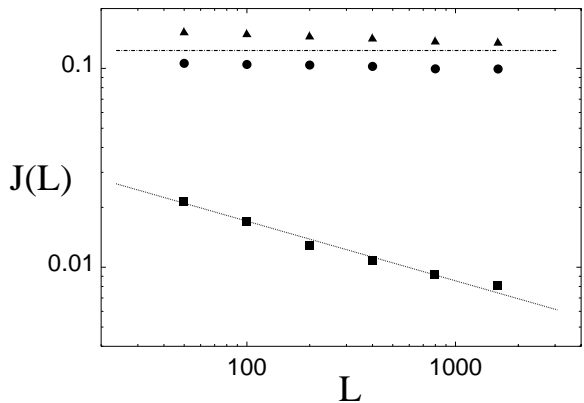,width=8cm}
\end{center}
\narrowtext
\noindent\caption{Variation of the steady-state current with the system size
for the three DASEP regimes of Fig. 1: (a) Homogeneous (circles), (b)
Segregated-Density (triangles) and (c) Vanishing-Current (squares).  In (a)
and (b), the current approaches a finite value in the thermodynamic limit
whereas in (c) the current vanishes as a power of the system size. The
dashed line corresponds to $J=0.125$ which is the limiting value of the
current in the regime (b) for the chosen values of the parameters.}
\label{fig:regimes2}
\end{figure}

      Examples of these behaviours are discussed in this paper for two
      types of lattice-gas models, namely the disordered drop-push process
      (DDPP) and the disordered asymmetric simple exclusion process
      (DASEP). The models are defined in detail in Sections \ref{sec:DDPP}
      and \ref{sec:DASEP} respectively, but for the purpose of discussion
      here, it suffices to note that the models are similar in that there
      is a maximum occupancy of each site in both, and are different in the
      dynamical moves ---  attempted nearest neighbour jumps in
      the DASEP, and slightly longer-ranged leapfrogging moves in the DDPP.

      The absence of detailed balance, together with the breaking of
      translational invariance, in disordered off-equilibrium systems makes
      the characterization of even the stationary state difficult in
      general.  It is shown that the steady state of the disordered
      drop-push process can be found explicitly -- the first such instance
      we are aware of, in a system with disorder, interactions and drive
      \cite{Krug_Evans}. This determination -- which is based on the
      condition of pairwise balance \cite{SRB} -- shows that a product
      measure form is valid in all dimensions. The form reflects the
      microscopic inhomogeneities coming from the underlying disorder, and
      results in a macroscopically homogeneous state.

      For the disordered asymmetric exclusion process, the steady state
      measure is not analytically characterizable, and we study the problem
      within a site-wise inhomogeneous mean-field theory and by numerical
      simulation.  The result depends crucially on whether or not the
      system has backbends, which are stretches of the lattice where the
      local bias is against the particle flow.  In the no-backbend case,
      when the average particle density is sufficiently away from 1/2, the
      spatial profile of the density has microscopic shocks, but is uniform
      on macroscopic scales (Fig.~\ref{fig:regimes1}a).  However, in a
      finite region around half-filling, disorder induces phase separation
      into macroscopic regions of high and low
      density (Fig.~\ref{fig:regimes1}b).  We give approximate arguments to
      understand the origin and nature of this phase separation, and to
      obtain the form of the phase diagram in the current-density plane.
      This sort of behaviour has also been seen earlier in a model with a
      single weak bond \cite{Janowsky}. We argue that disorder-induced
      phase separation is a generic feature of systems in which the current
      $J$ versus density $\rho$ shows a maximum at some intermediate
      density, in the absence of disorder.

      In the version of the DASEP in which the easy direction of hopping is
      itself a quenched random variable, the model represents a system of
      hard-core particles in a random potential with an overall downward
      tilt, but with backbends of arbitrary length.  Long backbends
      severely limit the maximum current that can flow through the system,
      and in fact the current decreases to zero as the system size
      increases (Fig.~\ref{fig:regimes2}); the system is in the
      vanishing-current regime.

      Although our emphasis in this paper is on the analysis of lattice
      models, we comment briefly on certain constraints that are important
      in a continuum description.  Such a description is expected to be
      valid for the large-scale, long-time behaviour, and is based on
      stochastic differential equations involving appropriate
      coarse-grained variables. It is argued that quenched randomness is
      manifest in random multiplicative coefficients in a gradient
      expansion.  Conservation of particle number -- which implies spatial
      constancy of the current in the steady state -- imposes strong
      constraints on these terms.

      In one dimension, using a well known mapping \cite{solids}, the
      particle models are equivalent to stochastic growth models of a 1-$d$
      interface moving in a 2-$d$ medium. The interface moves with a speed
      proportional to the current in the particle model. The
      disordered jump rates now become local growth rates which are
      disordered in a columnar fashion 
      for the moving interface \cite{columnar}. The three principal
      regimes of behaviour discussed above for the particle models
      translate into distinct regimes for interface motion, namely (i) a
      moving interface with normal roughness, (ii) a moving interface with
      large segments with different mean slopes, and (iii) an interface
      with different-slope segments, which is stationary in the
      thermodynamic limit.

      The paper is organized as follows. In Section \ref{sec:DDPP} we
      define and discuss the steady state properties of the disordered
      drop-push process in arbitrary dimensionality. The disordered
      asymmetric exclusion process with only forward-easy-direction of
      hopping, but quenched random rates, is discussed in Section
      \ref{sec:DASEP}; the case in which there are some
      backward-easy-direction bonds is discussed in Section
      \ref{sec:Sinai}. In Section \ref{sec:continuum} we discuss the
      constraints on a continuum description, while Section
      \ref{sec:height} discusses the implications of our results for models
      interface growth in the presence of columnar disorder. Section
      \ref{sec:conclusion} is the conclusion.

\section{Disordered Drop-Push Process : DDPP} \label{sec:DDPP}

\begin{figure}[tb]
\begin{center}
\leavevmode
\psfig{figure=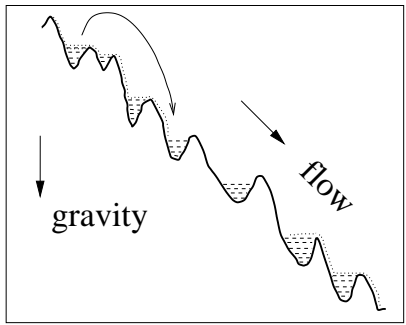,width=5.5cm}
\end{center}
\narrowtext
\noindent\caption{Schematic diagram of water flowing down a rugged
hill-side.  Water from a lake higher up cascades downhill, under the action
of gravity, until it finds a partially filled lake. The unequal capacities
of the lakes are the quenched variables in the system.}
\label{fig:lakes}
\end{figure}

\begin{figure}[tb]
\begin{center}
\leavevmode
\psfig{figure=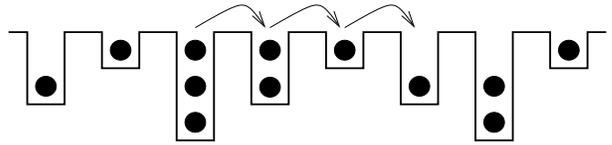,width=8cm}
\end{center}
\narrowtext
\caption{A Disordered Drop-Push Process (DDPP) configuration and move in 
$d=1$. }
\label{fig:ddpp-1d}
\end{figure}

    The drop-push process was initially introduced in \cite{BR,SRB} as a
    model of activated flow involving transport through a series of traps
    of equal depths. The dynamics consists of activated hops together with
    a cascade of overflows following each move.  The disordered version of
    the model may be considered as a discrete model of activated fluid flow
    down an inclined rugged slope with lakes of varying depths; see
    Figs.~\ref{fig:lakes},\ref{fig:ddpp-1d}. This is similar to
    above-threshold behaviour of the model considered in
    \cite{Narayan_Fisher}. In this section we show that the steady state
    and current can be found exactly in all dimensions for the DDPP and its
    generalizations.

   \subsection{The model}   \label{sub:defDDPP}

\begin{figure}[tb]
\begin{center}
\leavevmode
\psfig{figure=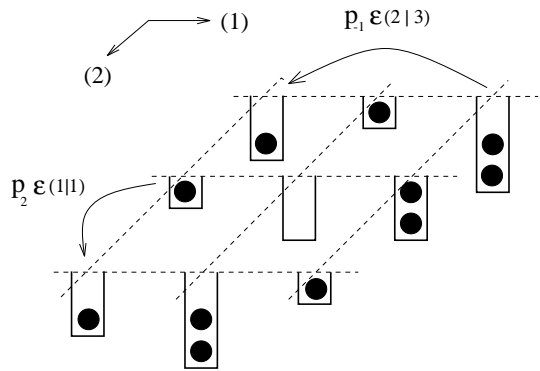,width=7cm,angle=-90}
\end{center}
\narrowtext
\caption{The DDPP model in $d=2$. The model can be generalized to $d>2$
(see text). The rates $\epsnl$ depend on the well depth $l_\r$ as well as
the occupation number $n_\r$.}
\label{fig:ddpp-2d}
\end{figure}

    The model in $d$-dimensions is defined on a hypercubic lattice with
    periodic boundary conditions along all the $d$ axes (with unit vectors
    \{$\enu |\nu=1\cdots d$\}). At each site $\r$ is a well which can hold
    at most $\lr$ particles (Figs.~\ref{fig:ddpp-1d},\ref{fig:ddpp-2d}) with
    $\lr$'s chosen independently from some probability distribution
    $P(l)$. The configuration $\cal{C}$ of the system is specified by
    specifying the set of occupation numbers \{$\nr$\} with ($0\le\nr\le
    \lr~, \forall\r$). Further, with each site $\r$ is assigned a set
    \{$\epsnl; \nr=1,\cdots,\lr$\} of positive random numbers chosen from
    some given distribution \cite{well_distbn}.  The dynamics is
    stochastic. In a time interval $dt$, with a probability
    $p_{\pm\nu}\epsnl dt$, the topmost particle in the well $\r$ hops out,
    and drops into well $\r\pm\enu$, i.e. into the adjacent well in the
    $\pm\nu$th direction. Here $\{p_{\pm\nu}; \nu=1,..,d\}$ are a set of
    site-independent positive numbers satisfying $\sum_{\nu=1}^d
    (p_\nu+p_{-\nu})=1$. Now, if well $\r\pm\enu$ is already full, then the
    particle gets pushed further {\it preserving the direction of the
    initial jump} to the next site and so on. The cascade of transfers
    terminates once a partially full well is encountered. Note that here
    the set of jump-rates \{$\epsnl$\} are site-dependent as well as
    functions of the occupation numbers. These rates, together with the
    well-depths \{$\lr$\}, constitute the quenched random variables in the
    model. The set of probabilities \{$p_{\pm\nu}$\} determines the
    direction of the global bias $\vE = \sum_{\nu=1}^d
    (p_\nu-p_{-\nu})\enu$ and, as will be shown in Section
    \ref{sub:imDDPP}, also the direction and magnitude of the steady state
    current in the model.  However, they do not enter the expression for
    the normalized invariant measure.

    Though all the results we will discuss holds for any arbitrary choice
    of the $\epsilon$'s, in a physical system they should be determined
    from the details of the trapping mechanisms etc., e.g. they may be
    taken to be of the Kramers form $\epsnl\propto exp[-g(\lr-\nr)]$ for
    situations where the jumps are activated \cite{kehr87}.

     \subsection{Invariant measure} 
     \label{sub:imDDPP} 

     The time evolution of the probability
     $\cal{P(\calC)}$ for the system to be in configuration $\calC$ is
     given by the master equation \cite{vanKamp}
\begin{equation} {d\over dt}{\cal{P({\cal C})}}
                    =\sum_{\cdprime}
W({\cdc}){\cal P}({\calC''})-\sum_{\cprime} W(\ccp){\cal P}({\cal C}).
\label{eq:master}
\end{equation}
      Here the $W$'s are the transition matrix elements identified with the
      rates $\epsilon$'s defined in the model. e.g. if the transition
      $\ccp$ involves moving the topmost of the $n_\r$ particles at $\r$ to
      $\rp$ along the $\nu$th direction, then
      $W({\ccp})=p_\nu\epsilon_\r(n_\r|l_\r)$. The steady-state or the
      invariant measure of the dynamics is the set of time independent
      weights $\{\mu({\cal C})\}$ satisfying (\ref{eq:master}) above.
      Hence the problem of finding the invariant measure reduces to that of
      finding a set of positive weights $\{\mu({\cal C})\}$ such that the
      total incoming flux into any configuration ${\cal C}$ (the first sum
      in (\ref{eq:master})) equals the total flux out of ${\cal C}$ (the
      second sum in (\ref{eq:master})).  The uniqueness of the invariant
      measure is ensured by the connectedness property of the $W$-matrix,
      i.e. every configuration can be reached from any other by a sequence
      of transitions \cite{vanKamp}.

      We claim that the (unnormalized) measure of configuration
      ${\cal C}(\{\nr\})$ in the steady state has the product form
\begin{equation} \mu({\cal C})=\prod_{\r} u_\r(\nr).
\label{eq:imDDPP}
\end{equation}
Here $u_\r$ are the single-site weights defined as
\begin{equation}
 u_\r(\nr) = \left\{ \begin{array}{ll}
               1 &\mbox{~~~if~$\nr=0$} \\
               \tau_\r(1)\cdots \tau_\r(\nr) &\mbox{~~~if~$0<\nr\le\lr$}
               \end{array} 
             \right. 
\end{equation}
where
$\tau_\r(\nr)=\epsilon_0/\epsnl$ with $\epsilon_0$ being a microscopic
rate.

       To show that (\ref{eq:imDDPP}) is indeed the invariant measure for
       the DDPP we show that it is possible to associate  
       configuration $\cdprime$ in one-to-one correspondence with every
       $\cprime$ obtained from $\cal {C}$ by an elementary transition such
       that
\begin{equation}
W({\ccp}) \mu({\cal C}) = W({\cdc}) \mu(\cdprime).
\label{eq:pwb}
\end{equation}
       The above is the condition of {\it pairwise balance} \cite{SRB}
       which ensures that the terms in the two sums on the right hand side
       of (\ref{eq:master}) cancel in pairs. Pairwise balance has been
       used earlier to find steady states of translationally invariant
       systems \cite{SRB,BPB}.  We now see that it can be used effectively
       to deduce the steady state of a disordered system as well.

\begin{figure}[tb]
\begin{center}
\leavevmode
\psfig{figure=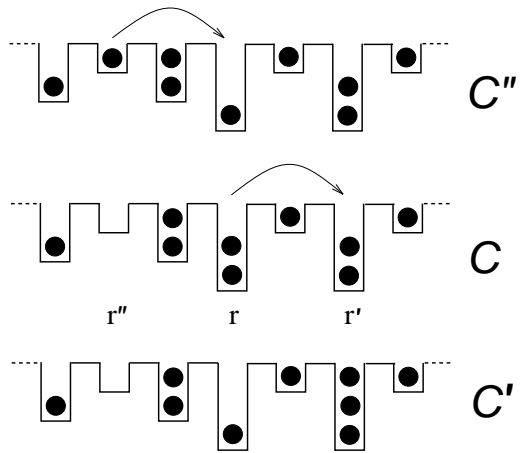,width=7cm,angle=-90}
\end{center}
\narrowtext
\caption{Construction of configurations satisfying the  pairwise balance
condition in $d=1$.}
\label{fig:pwb-2d}
\end{figure}

       Suppose the transition $\ccp$ involves hopping a particle at site
       $\r$ to a site $\rp=\r+\Delta r'\enu$ with all wells in between
       along the $\nu$th axis full (Fig.~\ref{fig:pwb-2d}). Also suppose
       the well $\rdp=\r-\Delta r''\enu$ is not full but all wells between
       $\rdp$ and $\r$ are full.  The configuration $\cdprime$ is
       constructed such that it is identical to $\calC$ at all sites except
       at the sites $\rdp$ and $\r$, at which $\ndp_\rdp=n_\rdp+1,~ \ndp_\r
       = n_\r-1$. With $\cdprime$ thus defined, the pairwise balance
       condition (\ref{eq:pwb})
\begin{equation}
\epsnl \prod_\r u_\r(n_\r) = \epsilon(\ndp_\rdp|l_\rdp)
                             \prod_\r u_\r(\ndp_\r)
\end{equation}
reduces to 
\begin{equation}
\epsnl u_{\rdp}(n_{\rdp})u_\r(n_\r)
       = \epsilon(\ndp_{\rdp}|l_{\rdp})
             u_{\rdp}(\ndp_{\rdp}) u_\r(\ndp_\r),
\label{eq:pwb-check}
\end{equation}
       since $\calC$ and $\cdprime$ differ only at the sites $\rdp$ and
       $\r$.  As can be explicitly checked, condition (\ref{eq:pwb-check})
       is satisfied in view of the form of the weights $u_\r(n_\r)$.  
       The prescription above ensures one-to-one correspondence between
       configurations $\cdprime$ and $\ccp$ transitions.

       In the limit of large $L$ we use the grand
       canonical description and write the normalized probability of
       configuration $\cal{C}$ in the steady state as
\begin{equation}
{\cal P}({\cal C}) = {z^{N_P}  \mu({\cal C}) \over {\prod_\r Z_\r}}
\end{equation}
where the site generating functions are given by 
\begin{equation}
Z_\r=\sum^{\lr}_{\nr=0}
     u_\r(\nr)z^{\nr}
\label{eq:sitegen-ddpp}
\end{equation}
     and $N_p$ is the number of particles in the configuration. Here $z$ is
     the fugacity and will be shown to be directly related to
     the steady-state current.  The mean particle density at site $\r$ is
     given as $\langle\nr\rangle = z \partial ln Z_\r/\partial z$
     \cite{angle_avg}.  Thus the fugacity $z$ is related to the mean
     particle density $\rho$ of the system through
\begin{equation}
\rho={1\over L^d}\sum_\r \langle\nr\rangle = {z\over V}\sum_\r {\partial
ln Z_\r \over \partial z }
\label{eq:implicit}
\end{equation}
     The steady state is characterized by a spatially uniform $z$ but
     inhomogeneous site densities.

     \subsection{Steady-state current}
     \label{sub:J0-DDPP}

      For simplicity of presentation, we first derive a closed form
      expression for the steady-state current $J_0$ for the 1-$d$ fully
      asymmetric model. Then we use these results to
      write down the current for the general $d$-dimensional case.

      (i) {\it $d=1$, with forward jumps}: In this case, for notational
            convenience, we replace the lattice site index $\r$ by a single
            integer index $i$. Further, we allow jumps only in one
            direction ($\nu=1: p_1=p, p_{-1}=0$). For an infinite system we
            write the current $J_{i,i+1}$ in the bond $(i,i+1)$ as
\begin{eqnarray}
   J_{i,i+1}& = &\sum_{j<i}\sum_{n_j=1}^{l_j}p\epsilon(n_j|l_j)P_j(n_j)
            \prod_{j<k\le i}P_k(l_k) \nonumber \\
           & & +\sum_{n_i=1}^{l_i}p\epsilon(n_i|l_i)P_i(n_i)
\label{eq:Ji-i1} 
\end{eqnarray}
            where 
\begin{equation}
P_i(n_i)=u_i(n_i) z^{n_i}/Z_i
\label{eq:site-prob}
\end{equation}
            is the probability that site $i$ has occupancy $n_i$. Note
            that the product form of the measure (\ref{eq:imDDPP}) has
            been exploited to write the above expression with
            decoupled joint probabilities.

            It is cumbersome, and not very instructive, to perform the
            sum in (\ref{eq:Ji-i1}) to obtain the current in a
            closed form.  Instead we use the spatial constancy of the
            current in the steady state to arrive at the result. We
            note that in the above expression the first term
            represents contribution of jumps originating from site $j$
            to the left of site $i$ with all sites in between
            full. The second sum represents jumps from the site $i$
            itself contributing to $J_{i,i+1}$. The first sum may be
            rewritten as

$ P_i(n_i)\left[\sum_{j<i-1}\sum_{n_j=1}^{l_j}\epsilon(n_j|l_j)P_j(n_j)
 \prod_{j<k\le i-1}P_k(l_k)\right. \\ \left.~~~~~~~~~~~~~
 +\sum_{n_{i-1}=1}^{l_{i-1}}\epsilon(n_{i-1}|l_{i-1})P_{i-1}(n_{i-1})\right]
 $.

 \noindent The quantity within square brackets [$\cdot$] is immediately recognised to
            be $J_{i-1,i}$ by comparing it with
            (\ref{eq:Ji-i1}). Physically, this means that all
            jumps contributing to $J_{i-1,i}$, for which site $i$ is
            completely full, also contribute to $J_{i,i+1}$. Hence we
            have the recursion
\begin{equation}
J_{i,i+1} = P_i(l_i)J_{i-1,i}
            +\sum_{n_i=1}^{l_i}p\epsilon(n_i|l_i)P_i(n_i)
\label{eq:Jrecur}
\end{equation}
            relating $J_{i,i+1}$ to $J_{i-1,i}$.
            Now, since $\{P_i(n_i)\}$ are the steady-state site
            probabilities, and since in the steady state all the bond
            currents must be equal (i.e. $J_{i,i+1}=J_{i-1,i}=
            ..... =J_0$), from (\ref{eq:Jrecur}) we obtain
\begin{eqnarray}
J_0 & = & {{\sum_{n_i=1}^{l_i}\epsilon(n_i|l_i)P_i(n_i)}
             \over{1-P_i(l_i)}} 
      =  p\epsilon_0 z
\label{eq:J0-1d}
\end{eqnarray}
      where (\ref{eq:imDDPP}) and (\ref{eq:site-prob}) have been used in
      the second step above. 

      Note that the steady-state current does not depend upon the detailed
      spatial arrangement of wells. It is only a function, through the fugacity
      $z$, of the density and the  total number of wells of different
      types in a particular realization of disorder.

      (ii) {\it $d=1$, with jumps in both directions: }
      We can write the current $J_{i,i+1}$ in bond $(i,i+1)$ as the
      difference between the current $J^r_{i,i+1}$ due to the rightward
      jumps and the current $J^l_{i,i+1}$ due to the leftward jumps. This
      can be done since in each cascade the direction of the initial jump
      is preserved. Now, we use the result for the fully asymmetric case
      above to each of these currents separately to obtain
      $J_{i,i+1} = J_0 = (p_1-p_{-1}) \epsilon_0 z$

     (iii) {\it $d>1$}: To generalize the above results to
     $d>1$ we note that for DDPP in $d>1$, the invariant measure
     (\ref{eq:imDDPP}) is the same if we single out a particular direction,
     say $\nu$, and allow jumps only along that direction. Together with
     the direction preservation of individual jumps, this allows us to
     write the expression for the current in any dimension:

\begin{eqnarray}
\vec J_0 &=& \left[\sum_{\nu=1}^d (p_\nu-p_{-\nu})\enu\right] \epsilon_0 z 
          = \epsilon_0 z \vE
\label{eq:J0-dd}
\end{eqnarray}
         where $\vE\equiv\sum_{\nu=1}^d (p_\nu-p_{-\nu})\enu$ is the external
         drive. As in the $d=1$ case, the magnitude and the direction of
         the steady-state current does not depend upon the detailed
         arrangements of the wells.

\subsection{Static two-point correlation functions}

          Because of the product form of the measure, the connected part of
          the equal-time density-density correlation function
\begin{equation}
G_\r(\Delta\r) = \langle n_\r n_{\r+\Delta\r}\rangle-\langle
n_\r\rangle\langle n_{\r+\Delta\r}\rangle
\label{eq:partcorr}
\end{equation}
          vanishes identically for $\Delta\r \ne 0$.  Consequently, the
          fluctuation of the number of particles in $r$ consecutive sites
          along a straight line can be computed exactly:
\begin{eqnarray}
\Gamma_i^2(r)& = & \left\langle\left[\sum_{j=i+1}^{i+r}\left(n_j-\langle
                  n_j\rangle\right)\right]^2 \right\rangle \nonumber\\ 
           & = & \sum_{j=i+1}^{i+r}
                 \left( z{\partial\over \partial z}\right)^2 \ln Z_j.
\label{eq:hhcor-DDPP}
\end{eqnarray}
\noindent The second step follows from the product form of the measure.

        For $d = 1$, a standard mapping discussed in Section VI introduces
        height variables defined by
\begin{equation}
h_i=\sum_{j\le i} 2(\langle n_j\rangle-n_j).
\label{eq:heightdef}
\end{equation}
\noindent Evidently, $\Gamma_i^2(r)$ is the equal-time height-height 
        correlation $\langle (h_{i+r}-h_i)^2\rangle$.  Averaging over the
        disorder distribution gives $\overline {\Gamma^2}(r)\sim r$
        implying that the `roughness' exponent $\alpha$ (defined by
        $\overline {\Gamma^2}(r) \sim r^{2\alpha}$) is $1/2$.
      
  \subsection{Two-rate DDPP model: Explicit results}
  \label{subsec:2rate}
    
     Let us consider a drop-push model where the maximum occupancy of each site
is restricted to one, i.e. $\lr=1, \forall\r$, but the hopping rates
$\epsilon_\r(1)$  are
disordered, and chosen independently from the binary distribution

\begin{equation}
Prob(x=\ea) = 1-f, ~~~~~~Prob(x=\eb)=f.
\end{equation}

\noindent  This model has the essential ingredients of disorder present in the
      original DDPP, yet it is simple enough that explicit, closed form
      relations between the mean density $\rho$ and fugacity $z$, and hence
      the steady-state current $\vec{J_0}$, can be written down.

      Let us denote by $Z_a$ and $Z_b$ the site generating functions
      for the $a$ and the $b$ sites respectively. Using
      (\ref{eq:sitegen-ddpp}) and (\ref{eq:imDDPP}), these are given by
      $Z_a=1+\epo z/\ea$ and $Z_b=1+\epo z/\eb$. Now, since the fractions of
$a$ and $b$ sites are $1-f$ and $f$ respectively, (\ref{eq:implicit})
reduces to 
\begin{equation}
\rho=(1-f){\epo z\over \ea+\epo z}+f{\epo z\over \eb+\epo z}.
\end{equation}
This can be easily
inverted to obtain $z$ as a function of $\rho$, e.g. for $f=1/2$ and
$\ea=\epo=\eb/q $ we obtain

\begin{equation}
z(\rho) = {\sqrt{(1-q)^2(1/2-\rho)^2+q}-(1+q)(1/2-\rho)\over 2(1-\rho)}
\label{eq:fugacity}
\end{equation}
Since $z(\rho)$ is known, the steady-state current is trivially obtained
from (\ref{eq:J0-dd}).

Finally, the correlation function $\Gamma_i^2(r)$ of (\ref{eq:hhcor-DDPP}),
upon disorder averaging, may be written as
\begin{eqnarray}
{\overline{\Gamma^2}(r)} & = & r\left( z{\partial\over \partial z}\right)^2
                              \left[(1-f)\ln Z_a + f \ln Z_b\right] \nonumber \\
                         & = &
                      {1\over 2}\left[{z\over(1+z)^2}+{qz\over(q+z)^2}\right] r
\end{eqnarray}
where $z$ is given by (\ref{eq:fugacity}) above.

  \subsection{Generalized Disordered Drop-push Process: GDDP}
  \label{subsec:GDDP}

    We may consider a generalized version of the drop-push process in
    which, in addition to the particle moves, independent hole moves are
    also allowed. For simplicity we restrict ourselves to the generalized
    version of the single occupancy DDPP introduced above. This generalized
    model may be regarded as the disordered lattice gas analogue of the
    Toom interface dynamics in the low-noise limit \cite{Toom}; see Section
    \ref{sec:height}. The techniques developed for the DDPP may be used to
    obtain the exact steady-state measure and other quantities such as
    current and static correlations provided a certain condition
    [(\ref{eq:constr-GDDP}) below] is met.

\begin{figure}[tb]
\begin{center}
\leavevmode
\psfig{figure=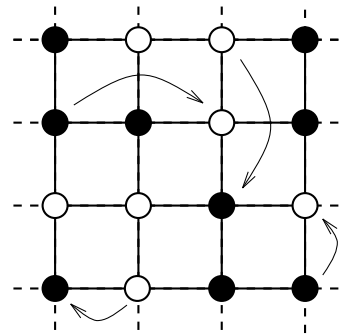,width=5cm,angle=-90}
\end{center}
\narrowtext
\caption{Generalized Disordered Drop-push Process (GDDP) configuration and
moves in $d=2$.}
\label{fig:gddp-model}
\end{figure}

    The model in $d$ dimensions is defined on a hypercubic lattice with
    periodic boundary conditions along all the $d$ axes (with unit vectors
    \{$\enu |\nu=1\cdots d$\}), similar to the DDPP.  Each site $\r$ of the
    lattice can hold either a particle ($\nr=1$) or a hole ($\nr=0$). The
    configuration $\cal{C}$ of the system is specified by specifying the
    occupation number of each well \{$\nr$\} with ($\nr\in\{0,1\},
    \forall\r$). Further, to each site $\r$ is assigned a pair of positive
    random numbers $(\alpha_\r,\beta_\r)$ chosen from some
    distribution. The dynamics is stochastic and is very similar to that
    for the DDPP dynamics: in a time interval $dt$, a particle at site $\r$
    is exchanged with the closest hole in the $\pm\nu$th direction with a
    probability $p_{\pm\nu}\ar dt$ (Fig.~\ref{fig:gddp-model}). For
    identical particles this move is equivalent to a cascade of particle
    moves terminating at the first vacant site as in the drop-push
    dynamics.  Likewise, in interval $dt$, a hole at site $\r$ is exchanged
    with the closest particle along the $\nu$th direction with probability
    $q_{\nu}\br dt$. This can be looked upon of as a cascade of hole-moves
    analogous to the cascade of particle moves.  Here, as in the DDPP, the
    $p_\nu$'s and $q_\nu$'s are all non-negative and satisfy
    $\sum_{\nu=1}^d (a_\nu+a_{-\nu})=1; a=p,q$. Further we chose $\alpha$'s
    and $\beta$'s such that
\begin{equation}
\ar\br= K,
\label{eq:constr-GDDP}
\end{equation} 
    \noindent where $K$ is a constant independent of $\r$.  As we will see
    below, this particular choice of the jump-rates allows the exact
    determination of the invariant measure of the model.  Physically this
    choice is quite reasonable since it implies that the sites which act as
    traps for particles (low $\ar$) are more transparent to holes (high
    $\br$) and vice versa. As pointed out earlier the non-disordered
    version of this model, i.e.  \{$\ar=\alpha, \br=\beta, \forall \r$\} in
    1-$d$, is the lattice gas equivalent of the low-noise driven Toom
    interface dynamics \cite{Toom}.

    The master equation (\ref{eq:master}) governing the time evolution of
    the system now includes terms corresponding to hole moves as well.
    Since each micro-step involves either only particle moves or hole
    moves, we use the principle of pairwise balance for the particle moves
    and hole moves separately.  We work in the grand canonical picture in
    the thermodynamic limit.

     If only particle moves were allowed the invariant measure would be
given by
\begin{equation}
\mu_{pcle}({\calC})=\prod_\r u_\r(n_r)
\label{eq:muC-p}
\end{equation}
     The single site weights $u_\r$ are given by
\begin{equation}
 u_\r(\nr) = \left\{ \begin{array}{ll}
               1 & \mbox{~~~if~$\nr=0$} \\ 
               \epsilon_0/\ar & \mbox{~~~if~$\nr=1$}
               \end{array} \right. 
\end{equation}
\noindent   Introducing the fugacity $z$ and site generating functions 
     $Z_\r=1+z u_\r(1)$ we can write the normalized single site
     probabilities as $P^{pcle}_\r(0)=1/Z_\r$ and
     $P^{pcle}_\r(1)=zu_\r(1)/Z_\r$.
 
     Similarly, with only the hole moves, the invariant measure has the product
form 
\begin{equation}
\mu_{hole}({\calC})=\prod_\r v_\r(n_r),
\label{eq:muC-h}
\end{equation}
where $n_\r=1$ ($0$) refers to the presence (absence) of a
     hole. The single site weights $v_\r$ are given by
\begin{equation}
 v_\r(\nr) = \left\{ \begin{array}{ll}
               1 & \mbox{~~~if~$\nr=0$} \\ 
               \epsilon_0/\br & \mbox{~~~if~$\nr=1$}
               \end{array} \right. .
\end{equation}
\noindent  Introducing the fugacity $y$ for holes and site generating 
     functions $Y_\r=1+y v_\r(1)$ we can write the normalized single-site
     probabilities as: $P^{hole}_\r(0)=1/Y_\r$ and
     $P^{hole}_\r(1)=y v_\r(1)/Y_\r$.

     Now, since each site is occupied either by a particle or a hole, we
     must have $P^{pcle}_\r(0)=P^{hole}_\r(1)$ and
     $P^{pcle}_\r(1)=P^{hole}_\r(0)$.  Using the detailed forms of
     $P^{pcle}_\r$'s and $P^{hole}_\r$'s, we arrive at the condition
     (\ref{eq:constr-GDDP}) with $K\equiv\epsilon_0^2 y z$.  If this
     condition is satisfied then the invariant measure for GDDP is given by
     either (\ref{eq:muC-p}) or (\ref{eq:muC-h}), since both are
     equivalent.

     In a similar manner as for the DDPP the current due to the particle
     moves and hole moves may be computed. The total particle
     current due to both types of moves, in $d$ dimensions, is given by
\begin{equation}
\vec {J_0} = \epsilon_0 z\sum_{\nu=1}^d (p_\nu-p_{-\nu})\enu 
             - \epsilon_0 y\sum_{\nu=1}^d (q_\nu-q_{-\nu})\enu
\end{equation}

    As for the DDPP, static density-density correlations in the steady
    state vanish identically on account of the product form of the
    steady-state measure.  In $d=1$ the height-height correlation is given
    by (\ref{eq:hhcor-DDPP}). 
 
\section{Disordered Asymmetric Simple Exclusion Process: DASEP \hfill} 
\label{sec:DASEP}
      The asymmetric simple exclusion process (ASEP) is a prototype model
      for studying nonequilibrium phenomena in the context of lattice gases
      \cite{ligget,spohn}. When discussing the effect of quenched disorder,
      it is important to distinguish between cases in which (a) the easy
      direction of hopping in each bond is the same but hopping rates are
      random, and (b) the easy direction is itself a random variable. The
      latter case is studied in Section \ref{sec:Sinai}. In this section,
      we consider a 1-$d$ system with disorder of type (a) and show that
      quenched disorder can induce macroscopic phase separation.  Using a
      variety of arguments we sketch the phase coexistence curve in the
      current ($J_0$) - mean density ($\rho$) plane. This agrees
      qualitatively with the results obtained from the Monte Carlo (MC)
      simulations.

      \subsection{The model} 
      \label{sub:defDASEP}
 
\begin{figure}[tb]
\begin{center}
\leavevmode
\psfig{figure=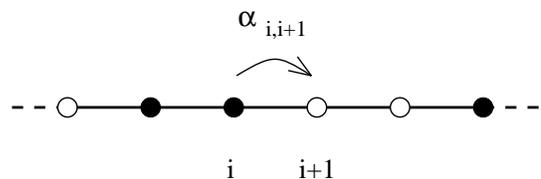,width=7cm,angle=-90}
\end{center}
\narrowtext
\caption{The disordered fully asymmetric simple exclusion process 
(DASEP) and moves in $d=1$.}
\label{fig:dasep-1d}
\end{figure}

      In one dimension, we define the disordered asymmetric simple
      exclusion process on a ring of $L$ sites. Each site can hold either
      $1$ or $0$ particle. To each bond $(i,i+1)$ of the lattice is
      assigned a quenched random rate $\alpha_{i,i+1}$ chosen independently
      from some probability distribution $Prob(\alpha)$. The dynamics is
      stochastic: in a time interval $dt$ a particle at site $i$ attempts
      to hop, with probability $p\alpha_{i,i+1} dt$, to site $i+1$. The
      move is completed if and only if site $i+1$ is unoccupied (see
      Fig.~\ref{fig:dasep-1d}).
      For the model defined above no analytically exact characterization of
      the steady-state measure could be obtained. A simpler 
      model in which there is only one defect bond has been studied in
      detail by Janowsky and Lebowitz \cite{Janowsky}, but in this case too
      the exact steady-state measure is not known.  We use Monte Carlo
      simulations and a mean-field approximation to demonstrate some
      striking effects of quenched disorder.

     \subsection{Current-density plot and density profile in steady
     state\hfill} \label{sub:imDASEP}

\begin{figure}[tb]
\begin{center}
\leavevmode
\psfig{figure=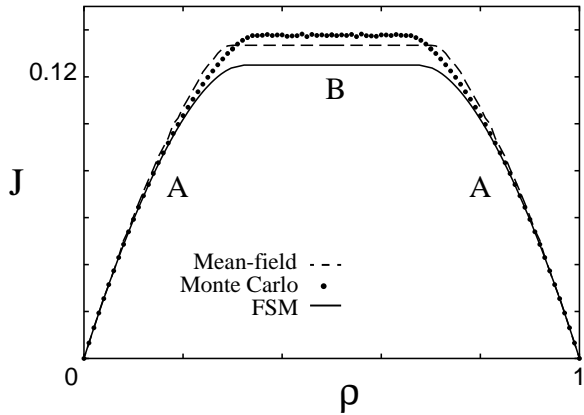,width=8cm}
\end{center}
\narrowtext
\caption{The current-density plot for the DASEP  for a given realization of
disorder for a system of size $L=8000$. The hopping rates are chosen from
the distribution ({\protect\ref{eq:binary}}) with $r=f=1/2$. The filled
circles are the MC results and the dashed line is the mean field curve. The
solid line represents the $J-\rho$ curve for the Fully Segregated Model
(FSM) for the same values of the parameters. }
\label{fig:dasep-jrho}
\end{figure}

     Figure \ref{fig:dasep-jrho} shows the steady-state current $J_0$ vs
     mean density $\rho$ plot, obtained from MC simulations, for
     a system of size $L=8000$ and the rates $\alpha$ chosen from the
     binary distribution
\begin{equation}
     Prob(\alpha=r)=f, ~~~~~Prob(\alpha=1)=1-f.
\label{eq:binary}
\end{equation}
\noindent Here $f$ is the fraction of weak bonds, and $r$ is a measure
     of the strength of the weak bonds. We used the values
     $r=1/2,f=1/2$ in our numerical work. For a specified mean density
     $\rho$, a random initial distribution of $N_p=\rho L$ particles
     on $L$ sites is chosen and the system is allowed to settle into a
     steady state by evolving it for a sufficiently large number of MC
     steps. Then the current across each bond is obtained by counting the
     total number of jumps across that bond, over a large number of MC
     steps. An average of all the bond currents thus computed is taken
     as $J_0$, as currents across all bonds are equal in the steady
     state. $J_0$ is a symmetric function of density around $\rho=1/2$
     as a result of a certain symmetry with respect to particle-hole
     interchange (see Appendix \ref{apx:PHsym}). As may be expected, the
     current for the disordered system lies between the corresponding
     values of the two pure reference systems with $r=1$ and $r=1/2$
     on all the bonds.  The more striking qualitative difference
     between the disordered and pure systems is the appearance of a
     plateau (Regime B in Fig.~\ref{fig:dasep-jrho}) for a range of
     densities $|\rho-1/2|\le \Delta$. In this regime, the current is
     independent of the mean density and equals the maximum allowed
     current in the system.  The approximate size $\Delta$ of regime
     B, which is a function of $r$ and $f$, is obtained in subsection
     \ref{sub:FSM} below.

\begin{figure}[tb]
\begin{center}
\leavevmode
\psfig{figure=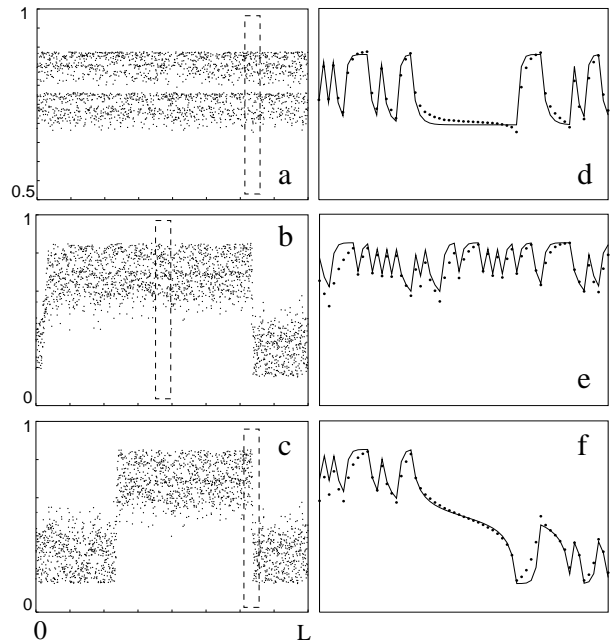,width=8cm}
\end{center}
\narrowtext
\caption{Density profiles for the DASEP for a system of size $L=8000$ for a
given realization of disorder at three fillings (a) $\rho=0.8$, (b)
$\rho=0.6$, and (c) $\rho=0.5$. In d,e and f are shown the blow-ups of the
regions enclosed in the dashed boxes in a,b and c respectively. Circles are
MC profiles and the continuous lines are mean-field results.}
\label{fig:dasep-profiles}
\end{figure}

     We studied the steady state density profiles characterized by the set
     of site densities $\{\rho_i\equiv\langle n_i \rangle\}$ in both
     regimes A and B, using MC simulations.  We found that in regime A the
     system is homogeneous on a macroscopic scale, while in regime B it
     shows macroscopic density segregation.  Figure
     \ref{fig:dasep-profiles} shows the steady state density profiles for
     three representative mean densities --- one from regime A and two from
     regime B. Evidently there is a large qualitative difference between
     the profiles in the two regimes.  In regime A
     (Fig.~\ref{fig:dasep-profiles}a), there are density variations
     (shocks) only over microscopic scales; coarse-graining over a few
     lattice spacings leads to a spatially uniform density.  In contrast to
     this, in the profiles corresponding to regime B
     (Fig.~\ref{fig:dasep-profiles}b), there are density inhomogeneities
     over length scales comparable to the system size $L$, in addition to
     the shocks on a microscopic scale.  This segregation into high and low
     density phases, with large shocks separating them, is reminiscent of
     phase separation, and occurs over the full range of mean particle
     densities corresponding to regime B.  A qualitatively similar
     phenomenon has been found in a system with one defect bond, studied in
     \cite{Janowsky}.

     The {\it number} of different large-scale regions of high and low
     density shows fluctuations from one realization of disorder to
     another.  As the size of the system is increased, we monitored the
     mean number of these regions, and found that it is nearly constant, or
     perhaps increases very mildly -- certainly, much less fast than
     linearly in the size of the system.  This implies that the
     characteristic length scale of density segregation increases
     indefinitely in the thermodynamic limit.

     \subsection{ Mean-field approximation \hfill}
     \label{sub:MFA}
      
      We now turn to a mean-field approximation, which assumes no
     correlations between site densities in the steady state.  We will see
     that it captures most of the steady-state features found in the MC
     simulations above, including the steady-state density profile.
     
     The time-averaged steady-state current $J_{i,i+1}$ in the bond
     $(i,i+1)$ is given by $J_{i,i+1} = \alpha_{i,i+1} \langle
     n_i~(1-n_{i+1})\rangle $.  In view of the mean-field approximation
     $\langle n_i n_j\rangle =\langle n_i\rangle\langle n_j\rangle$ this
     reduces to
\begin{equation} J_{i,i+1} = \alpha_{i,i+1}\langle
     n_i\rangle~\langle 1-n_{i+1}\rangle \label{eq:MFA} 
\end{equation}

     To compute the steady-state current $J_0$ as a function of the mean
     particle density $\rho$ for a given realization of disorder, we use
     the two iteration schemes described below, which yield equivalent
     results.

     (i) {\it Constant current iteration scheme.} For a given system
     of size $L$ and for a fixed value $J_0$ for the current, we
     iterate the following set of equations
\begin{equation}
    \rho_{i+1}=1-J_0/(\alpha_{i,i+1}\rho_i) ~~~~i=1,..,L
\end{equation} 
     around the chain starting with, say, some value $\rho_1$ (periodic
     boundary conditions imply $\rho_{i+L}=\rho_i$). If $J_0$ is less than a
     certain value $J^{MF}_{max}$, which is the maximum current supported
     by the system within the mean-field approximation, then the iteration
     scheme converges, i.e. we get all the site densities in the physically
     acceptable range $[0,1]$. The average of these site densities gives
     the mean density of the system corresponding to the stationary current
     $J_0$. There are in general two values of the mean particle density
     corresponding to an allowed value of $J_0$ and the
     iteration scheme converges to one or the other depending on the
     initial value of the density $\rho_1$.
  
     However, in this scheme we find that the number of iterations required
     for convergence increases without bound, as the trial value $J_0$ gets
     closer to $J^{MF}_{max}$ from below.  This divergence of the iteration
     scheme is presumably due to the existence of the plateau in the
     $J~vs~\rho$ curve, i.e. there exist many values of $\rho$ for $J_0$
     very close to $J^{MF}_{max}$. Hence to obtain the density profile for
     $\rho$ in the density segregated regime we resort to the {\it constant
     density} iteration scheme described below.

     (ii) {\it Constant density iteration scheme.}  In this scheme we begin
     by assigning site densities $\{0\le\rho_i(0)\le 1\}$ to the lattice
     sites subjected to the constraint ${1\over L}\sum_i \rho_i(0)
     =\rho$. The site densities are updated in parallel according to:
\begin{equation}
     \rho_i(t+1)=\rho_i(t)+J_{i-1,i}(t)-J_{i,i+1}(t);~~~i=1,..,L 
\end{equation}
     where $J_{i,i+1}(t)=\al_{i,i+1}\rho_i(t)[1-\rho_{i+1}(t)]$ in view of
     (\ref{eq:MFA}).

     We refer to this as the constant density scheme, since in each
     iteration the total density remains unchanged, i.e
     $\sum_i\rho_i(t+1)=\sum_i\rho_i(t)$. After a sufficient number of
     iterations, which depends upon the starting mean density $\rho$,
     the set of site densities converge to a set of numbers
     \{$\rho_i$\} and the current on each bond converges to the steady
     state current $J_0$.

     The steady-state density profiles and the $J_0~vs ~\rho$
     plot ($0\le\rho\le1$) for a given configuration of disorder obtained
     using these schemes is shown in Figs.~\ref{fig:dasep-profiles} and
     \ref{fig:dasep-jrho} respectively.  It is evident that the mean-field
     approximation (\ref{eq:MFA}) reproduces quite well not only the
     $J-\rho$ relationship, but also the density profile, including the
     locations of shocks, though not the shapes of individual shocks.

     \subsection{Qualitative explanation of phase separation}
     \label{sub:MF-gross}    

     Although the mean-field approximation of the previous subsection
     successfully reproduces many features in the steady state, it does not
     yield a simple understanding of the phase separation
     (Fig.~\ref{fig:dasep-profiles}) or the plateau in the $J$--$\rho$
     curve (Fig.~\ref{fig:dasep-jrho}) in terms of the macroscopic
     parameters of the model.  We conjecture that underlying the behaviour
     of the DASEP in different regimes is a Maximum Current principle: For
     a given mean density, the system settles into a steady-state 
     which maximizes the stationary current. Such a maximum
     current principle has been used to describe phase separation in the
     asymmetric simple exclusion process with open boundary conditions in
     \cite{Krug91}.
 
\begin{figure}[tb]
\begin{center}
\leavevmode
\psfig{figure=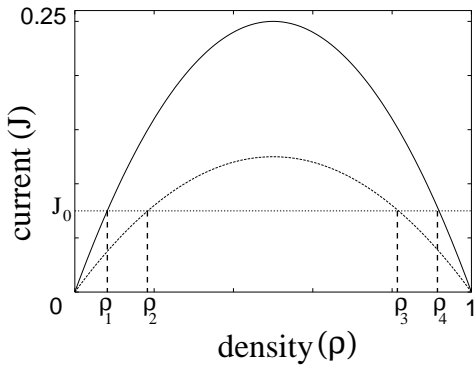,width=6.5cm}
\end{center}
\narrowtext
\caption{Origin of phase separation in DASEP. The two parabolas
$J=r\rho(1-\rho);~r=1,1/2$, represent the
$J-\rho$ plots for the two reference non-disordered systems.}
\label{fig:phases-expl}.
\end{figure}

     To use the maximum-current principle to have a qualitative
     understanding of the phase separation in DASEP, let us assume that the
     density in each stretch of like bonds is uniform. This approximation
     is in fact exact in the Fully Segregated Model (FSM) discussed in the
     following subsection. Let us denote stretches of $\alpha=1$ bonds by X
     and stretches of $\alpha=r<1$ bonds by Y. The two parabolas in
     Fig.~\ref{fig:phases-expl} are the steady-state $J~vs~\rho$ curves for
     the two pure reference systems all X and all Y. In the disordered
     system, since the steady-state current has to be spatially constant,
     the possible densities are given by the four intersections of the line
     $J=J_0$ with the two parabolas. If the mean density is in the range
     $\rho\le 1/2-\Delta$ (or $\rho\ge 1/2+\Delta$), then the allowed
     densities for the X and Y stretches are $\rho_1,\rho_2$ (or
     $\rho_4,\rho_3$) respectively. The current is in fact determined by
     the density constraint
     $(1-f)\rho_{1,4}(J_0)+f\rho_{2,3}(J_0)=\rho$. The variation of density
     between $\rho_1$ and $\rho_2$ (or $\rho_3$ and $\rho_4$) between X and
     Y stretches corresponds to the `sub-bands' seen in
     Fig.~(\ref{fig:dasep-profiles}a). On a macroscopic scale, however, the
     system has uniform density. Now consider what happens when the mean
     density is brought closer to $1/2$. From Fig.~\ref{fig:phases-expl},
     it is evident that the current would tend to increase, and would
     eventually reach the maximum allowed value $J_{max}^Y$ (which equals
     $1/4$ in the thermodynamic limit, as argued below). As the density is
     increased further, the current remains constant at $J_{max}^Y$, in
     accordance with the maximum current principle, and the excess density
     is taken care of by converting some of the X stretches of density
     $\rho_1$ into ones with $\rho_4$ (or vice versa if $\rho > 1/2$). This
     conversion takes place adjacent to the largest stretch of Y bonds,
     leading to two macroscopic regions of different mean densities -- one
     with densities $(\rho_1, \rho_2)$ for the X and Y stretches, and the
     other with $(\rho_4, \rho_3)$. The position of the principal shock
     separating these regions is at the location of the largest Y
     stretch. In the DASEP, the assumption of uniform density in each
     stretch is not really true, on account of the finite length of most of
     the stretches. However, the above argument provides a qualitative
     picture towards understanding the reason for phase separation in the
     DASEP.
     In a certain respect, the reason for the phase separation in the DASEP
     is similar to that in the single defect bond model studied in
     \cite{Janowsky} --- both have local segments in the system where the
     maximum allowed current is less than that allowed everywhere else. In
     the single defect bond model, phase separation takes place when the
     current carried by the rest of the system, with presumed uniform
     density, is larger than the maximum current allowed through the weak
     bond. In the DASEP, with an extensive number of weak bonds, the {\it
     largest stretch} of weak bonds acts as the current-limiting
     segment. The length of this stretch increases as $\ln L$ with system
     size, and in the thermodynamic limit the maximum allowed current in
     this stretch is ${1\over 4}r$ -- equal to the maximal current in a
     pure system with only weak bonds.

     The essential point leading to phase separation is thus the maximum
     current principle, coupled with localized current-limiting regions in
     the system. In the DASEP, this limit is determined by long stretches
     of weak bonds, and similar considerations should apply in related
     models.  Consequently, we would expect a density-segregated phase in
     disordered versions of models which, in the absence of disorder,
     display a maximum in the steady-state current as a function of
     density.
 
     \subsection{The Fully Segregated Model} \label{sub:FSM}

     It is useful to define a model for which many of the approximations
     made in the previous subsection are actually exact. To this end, we
     study a Fully Segregated Model (FSM), which is obtained from the
     binary random model above by arranging all like bonds together. Thus,
     in this model, one has {\it two} large stretches of X and Y, of
     lengths $(1-f)L$ and $fL$ respectively. For the FSM in the
     thermodynamic limit, the assumption of uniform density within each
     stretch is justified, as correlations due to the junctions decay with
     increasing separation, and may be neglected in the bulk \cite{FSM}.

\begin{figure}[tb]
\begin{center}
\leavevmode
\psfig{figure=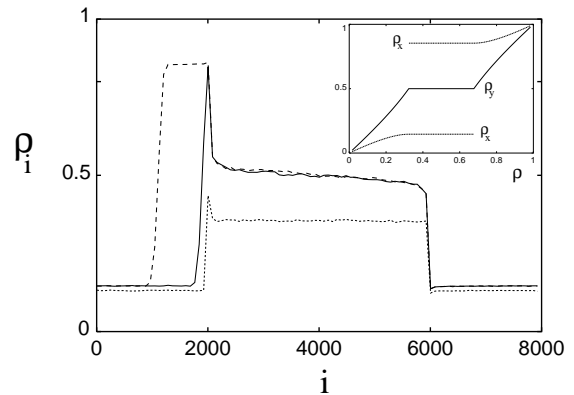,width=7.5cm}
\end{center}
\narrowtext
\caption{Representative density profiles of the FSM with $r=f=1/2$ at 
three different fillings: (a) $\rho=0.24$ (dotted) (b) $\rho=\rho_c^-
=0.324$ (solid) and (c) $\rho=0.4$ (dashed). The Y stretch ($r=1/2$) is in
the range $i\in[2000,6000]$. The inset shows the variation of the bulk
densities in the X and Y stretches as a function of the filling
$\rho$.}
\label{fig:dp-seg}
\end{figure}

     Steady-state MC density profiles for the FSM at three different
     fillings $\rho\le 1/2$ are shown in Fig.~\ref{fig:dp-seg} ---
     symmetry under particle-hole exchange implies analogous behaviour
     for $\rho\ge 1/2$ (Appendix \ref{apx:PHsym}). For low densities
     ($\rho<\rho_c^-$), the two stretches have uniform bulk
     densities $\rho_x$ and $\rho_y$ related to each other by the
     requirement of equality of the two bulk currents,
\begin{equation}
\rx(1-\rx)=r\ry(1-\ry)=J_0,
\label{eq:J-xy}
\end{equation}
and the density constraint 
\begin{equation}
(1-f)\rx+f\ry=\rho. 
\label{eq:contr-rho}
\end{equation}
These three equations determine $\rx, \ry$ and $J_0$ uniquely for a given
$\rho$. For $f=1/2$ we obtain
\begin{eqnarray}
\ry &=& {4\rho-1-r\pm\sqrt{(4\rho-1-r)^2+8(1-r)\rho(1-2\rho)}\over
{2(1-r)}}\nonumber \\ 
\rx&=&2\rho-\ry, ~~J_0=\rx(1-\rx)
\end{eqnarray}
     This is analogous to the macroscopically homogeneous state of the
     fully random system.
     As the mean density is increased, the density of each stretch
     increases, until, at a critical density $\rho=\rho_c^-={1\over
     2}-{1\over 4}\sqrt{1-r}$ (the corresponding critical density on the
     higher side is $\rho_c^+=1-\rho_c^-$), $\ry$ equals $1/2$
     (Fig.~\ref{fig:dp-seg}).  At this density, the current equals the
     maximum possible current in Y, namely, $J_0=J_{max}=r/4$.  As $\rho$
     is increased further, $\ry$ and $J_0$ remain constant at $1/2$ and
     $r/4$ respectively. The density change is adjusted by creating a
     density inhomogeneity in stretch X. The two densities $\rx^{high}$ and
     $\rx^{low}$ are related by $\rx^{high}+\rx^{low}=1$ so that the
     currents in the two phases are equal,
     i.e. $\rx^{high}(1-\rx^{high})=\rx^{low}(1-\rx^{low})=r/4$. This
     implies $\rx^{high,low}=(1\pm\sqrt{1-r})/2$ (see inset in
     Fig.~\ref{fig:dp-seg}).  The fraction of these phases can be determined
     from a lever rule and are given by
     $f^{high,low}=|\rho-\rho^\mp_c|/(\rx^{high}-\rx^{low})$. This locking
     of the density in the Y stretch at $\ry=1/2$ is a direct consequence
     of the maximum current principle introduced above: any change of $\ry$
     from $1/2$ would reduce the current in Y, and hence in the full
     system.  All the arguments above can be applied for $\rho>1/2$ because
     of particle-hole symmetry. Thus for $\rho^-_c<\rho<\rho^+_c$ the state
     of the segregated model is analogous to the phase separated regime B
     of the random model. The size of regime B in the FSM is given by
     $2\Delta=\rho_c^+-\rho_c^-=\sqrt{1-r}/2$. It closely approximates the
     size of the B regime in the DASEP (Fig.~\ref{fig:dasep-jrho}).

\subsection{Phase-coexistence curve}
     \label{sub:pcc}

\begin{figure}[tb]
\begin{center}
\leavevmode
\psfig{figure=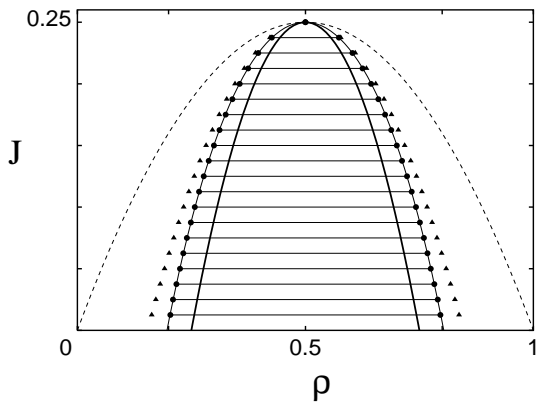,width=7.6cm}
\end{center}
\narrowtext
\caption{The phase coexistence curve for the FSM and the DASEP for $f=1/2$.
The solid parabola is the coexistence curve of the FSM. The circles and the
triangles are respectively the MC and mean-field phase-coexistence curves
for the DASEP. The dashed parabola $J=\rho(1-\rho)$ demarcates the allowed
region for the DASEP. }
\label{fig:phase-diag}
\end{figure}

     For the FSM, as $r$ is varied we obtain
     different $J_0~vs~\rho$ curves. The phase-coexistence curve in the
     current-density plane in the parametric form
\begin{equation}
J_c={r\over 4}, ~~~~~\rho_c={1\over 2}\pm{1\over 4}\sqrt{1-r}
\label{eq:ph-diag}
\end{equation}
    which is the parabola $J_c=1/4-(1-2\rho_c)^2$ in
    Fig.~\ref{fig:phase-diag}.

    The difference between the phase boundaries of the DASEP and FSM
    Fig.~\ref{fig:phase-diag} comes from the fact that the interspersed
    weak-bond stretches in the DASEP have finite lengths, and the density
    in these small stretches is not quite equal to $1/2$.  Close to the
    phase separation, we anticipate the mean density in a Y-stretch of
    length $l$ in the DASEP to be of the form $\rho_Y(l)=1/2\pm
    A(r)/l^\alpha(r)$, with $\alpha(r)>0$.  Using the distribution of the
    $Y$ stretches, namely $P_Y(l)=2^{-l}$, we obtain the $r$ dependence of
    the critical density
\begin{equation}
 \rho_c^\pm = {1\over 4} + \rho_X \pm A(r) \sum_l P_Y(l) \rho_Y(l) $$
\end{equation}
    \noindent where $\rho_X(r<<1)\sim r$ is the density in the X stretches
    near the phase transition.  Comparing with the phase diagram for the
    FSM in Fig.~\ref{fig:phase-diag}, the correction term $A(r)$ seems to
    be positive for all $r$. Further, let us suppose that the current in
    the FSM is a lower bound to that in the DASEP, as suggested by
    Fig.~\ref{fig:phase-diag}.  Then the coexistence curve for the random
    system must be quadratic near the critical point ($J^0=1/4,
    \rho^0=1/2$) -- being bounded by two quadratics, namely, the $J-\rho$
    curve for the pure system $J=\rho(1-\rho)$, and the coexistence curve
    of the FSM $J=1/4-(1-2\rho)^2$.

     \subsection{Correlations in the steady state}
     
\begin{figure}[tb]
\begin{center}
\leavevmode
\psfig{figure=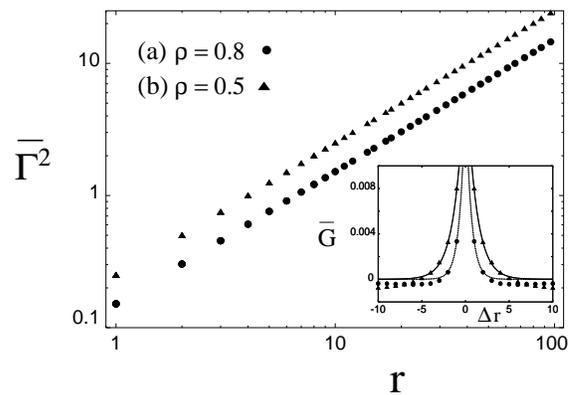,width=7.6cm}
\end{center}
\narrowtext
\caption{ Height-height correlation function $\overline{\Gamma^2}(r)$
for DASEP: (a) in Homogeneous (circles), and (b) Segregated-Density
(triangles) regimes. Inset shows the site averaged density-density
correlation function $\overline G(\Delta\r)$ defined in
({\protect\ref{eq:partcorr}}). The small negative values at large
$|\Delta\r|$ arise due to the finite size of the system.}
\label{fig:dasep-corln}
\end{figure}

     Figure \ref{fig:dasep-corln} shows the Monte Carlo results for the
     site-averaged density-density and height-height correlation functions
     $\overline{G}(\Delta\r)~{\rm and}~ \overline{\Gamma^2}(r)$ in both the
     homogeneous and density-segregated regimes of the DASEP.
     $\overline{G}(\Delta\r)$ is seen to decay rapidly over a few lattice
     spacings, accounting for the success of the mean field approximation.
     It is found that $\overline{\Gamma^2}(r)$ grows as $r$ --- implying a
     roughness exponent $\alpha=1/2$.

\section{DASEP with  backbends}
\label{sec:Sinai}

     As discussed at the beginning of Section \ref{sec:DASEP}, the
     introduction of randomness in the easy direction of individual bonds
     alters the properties of one-dimensional disordered exclusion process
     in a crucial way.  We study this in this section.

     The model is defined as follows: Assign quenched arrows (pointing
     either right or left) independently to each bond of a periodic chain,
     with probability $f < {1\over 2}$ for left arrows, and $1-f$ for right
     arrows.  An arrow defines the easy direction of hopping on each bond:
     a particle-hole exchange across a bond occurs with rate $w (1+g)$ if
     the particle moves along the direction of the arrow, and $w (1-g)$ if
     it moves opposite to the arrow.  Since $f < {1\over 2}$, there is an
     overall tendency for particles to circulate rightward, and the
     question is whether there is a nonzero current even in the
     thermodynamic limit.

     The model represents a system of hard-core particles in a random
     potential with a downward tilt.  A conglomeration of left pointing
     arrows constitutes a backbend, where the potential climbs up before
     going down again.  Within mean-field theory it is possible to obtain
     an upper bound $J_\l$ on the current that can be carried by mutually
     excluding particles through a backbend of length $\l$ \cite{RB}.  To
     this end, consider biased diffusion of hard core particles in the
     segment [$0, l$] of a 1-d lattice, with the `optimal' boundary
     conditions $\rho(0) = 1$ and $\rho (l) = 0$; these boundary conditions
     force the largest possible current through the segment, opposite to
     the bias.  The master equation that describes transport is invariant
     under interchange of particles and holes and simultaneous relabelling
     of sites in reverse order, i.e. $n_j \rightarrow \bar n_{l - j}$ The
     boundary conditions respect this symmetry, implying that the
     steady-state density $\rho(j)$ at site $j$ satisfies $\rho(j) = 1 -
     \rho (l - j)$.  Thus in the steady state the number of particles in
     the backbend is $l/2$ irrespective of the strength of the bias $g$.
     The principal effect of increasing $g$ is to sharpen the region which
     marks the transition from the particle-rich half of the backbend to
     the hole-rich half.  The steady-state profile approaches a step
     function centered at $j = l/2$ as $ g \rightarrow 1$.

     The current in the steady state is the number of particles crossing
     site $l$ in unit time. Results of a Monte Carlo study \cite{RB} are
     consistent with the large-$l$ asymptotic behaviour 
\begin{equation}
 J_l \sim \exp  \left( - {1 \over 2}~l/\Lambda(g) \right).
\label{eq:s1}
\end{equation}
     where $\Lambda(g)$ is a bias-induced length given by $\Lambda^{-1}(g)=
     \ln \{(1+g)/(1-g)\}$.  This can be seen by writing the current within
     a mean field approximation as $J = W(1 + g) \rho_j (1 - \rho_{j+1}) -
     W(1 -g) \rho_{j+1} (1 - \rho_j)$, and finding the value of $J$ for
     which the boundary conditions $\rho_0 = 1$, $\rho_l = 0$ hold.  For
     $l>>\Lambda(g)>>1$,this leads to $J \approx 2g~e^{-lg}$ \cite{RB}, in
     agreement with (\ref{eq:s1}) when $g$ is small.

     The origin of the factor $1\over2$ in the exponent in (\ref{eq:s1})
     has been discussed in \cite{RB}, and we recount the argument in brief.
     The transport of a single particle through the backbend involves two
     (approximately) causally independent steps which occur in parallel:
     (i) the topmost particle (located at site $k \approx l/2$ in large
     fields) has to be activated a distance $l/2$, which requires an
     activation time $\tau_{1/2} \sim \exp \left({1\over 2}~l /\Lambda(g)
     \right)$ and, (ii) the consequent hole that remains in the
     steady-state distribution moves to the bottom and is filled up, by
     moving each of $l/2$ particles up through a lattice spacing.  The time
     required is $\tau_{1/2}$ again.  The current $J$ is thus proportional
     to $\tau^{-1}_{1/2}$, and consequently follows (\ref{eq:s1}).

\begin{figure}[tb]
\begin{center}
\leavevmode
\psfig{figure=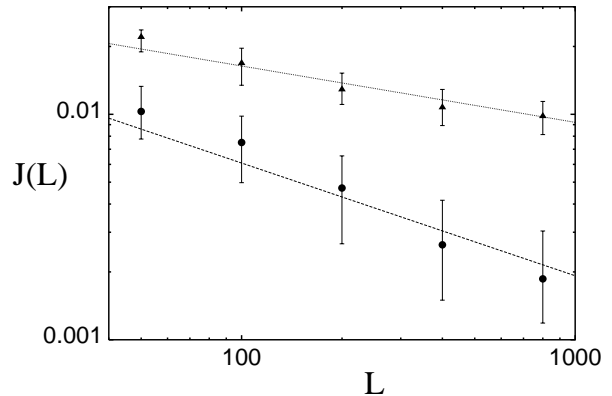,width=8cm}
\end{center}
\narrowtext
\caption{Size dependence of the steady-state current in the backbend
model (DASEP with some reversed bonds) for two sets of parameters: (a)
$g=0.33, f=0.25$ (triangles) and (b) $r=0.54, f=0.3$ (circles). Each point
represents an average over $40$ realizations of disorder in (a) and $100$
realizations in (b). The straight lines have slopes of $-\theta/2$ with
$\theta=0.5$ and $1.0$ respectively as predicted by
({\protect\ref{eq:s2}}).}
\label{fig:JL_sinai}
\end{figure}

     Since, for fixed $g$, the largest current that can flow through a
     long backbend ($l>>\Lambda(g)$) is exponentially small in its
     length, the largest current through the 1-d lattice of length $L$
     is determined by the length $\l^*(L)$ of the largest backbend
     encountered.  Since the probability of occurrences of $\l$
     consecutive left-pointing arrows on bonds is proportional to
     $f^\l$, we may estimate $\l^*$ from $Lf^{\l^*} = C$ where $C$ is
     a constant of order unity.  Substituting in (\ref{eq:s1}), we
     find that the current falls with increasing lattice size as
\begin{equation}
J(L) \sim L^{-{1\over2} \theta} 
\label{eq:s2}
\end{equation}
     with $ \theta^{-1} = \Lambda(g)\ln f$.
     Thus the current is expected to decay as a power law in $L$, with a
     bias-dependent power, and to vanish in the thermodynamic limit.
     Figure \ref{fig:JL_sinai} shows the result of Monte Carlo simulations.

     As with the milder sort of disorder discussed in Section III, the
     state is strongly inhomogeneous and shows macroscopic regions of high
     and low density. Figure \ref{fig:regimes1}c shows the time-averaged
     density profile for a typical configuration of bonds. There is a large
     shock around the rate-limiting backbend, which separates the two
     regions.

     For fixed lattice size $L$ with an associated longest backbend
     $l^*(L)$, the current is a nonmonotonic function of $g$. This can
     be seen as follows. If $g$ is low enough that
     $\Lambda(g)>>l^*(L)$, linear response theory would imply that
     current $J$ grows linearly with $g$. On the other hand, if $g$ is
     large enough that $\Lambda(g)<<l^*(L)$, the current falls with
     increasing $g$ according to (\ref{eq:s1}). In between, $J$
     achieves a local maximum when $\Lambda(g=g_{max})\simeq l^*(L)$,
     which implies $g_{max}\sim 1/\ln L$.

     The argument given above implies that the current carried by a system
     of hard-core particles through the randomly backbending lattice
     vanishes in the thermodynamic limit, no matter how small the bias
     $g$. This is in contrast to the behaviour of {\it noninteracting}
     particles in the same random environment, where the drift velocity
     vanishes only if the bias is strong enough \cite{BD,BR}. The
     difference can be traced to the possibility, in the noninteracting
     case, of compensation by a large build-up of density at the bottom of
     a backbend, which then succeeds in driving a finite current over the
     backbend.  This option does not exist once repulsive hard-core
     interactions come into play, and the current vanishes in the
     thermodynamic limit.

\section{Continuum description}
\label{sec:continuum}

     It is interesting to ask whether the behaviours found above in the
     disordered lattice gases can be reproduced using a continuum
     description of the problem. Though we have not pursued this question
     to its logical end, we discuss in this section some general
     constraints that a continuum description should satisfy.

      The steady-state of the disordered system of interacting particles is
     described by a spatially varying time-averaged density profile
     $\rho_0(\r) \equiv \langle n(\r) \rangle$.  The time evolutions of
     fluctuations around the mean density profile are governed by the
     continuity equation
\begin{equation}
{\partial\over\partial t} \rho(\r,t) + \nabla\cdot{\vec J}(\rho,\r,t) = 0.
\label{eq:continuity}
\end{equation}
     The coarse-grained current ${\vec J}$ may be phenomenologically
divided into three parts
\begin{equation}
{\vec J}(\r,t)=\vec{J}_{sys}(\rho(\r,t),\r)+ {\vec J}_{diff}
 +\vec{\epsilon}(\r,t).
\label{eq:J=j+j+j}
\end{equation}
     A local hydrodynamic assumption has been made in writing the {\it
     systematic} part of the current, $\vec{J}_{sys}$ as a function of the
     local density, $\rho(\r,t)$. The explicit $\r$ dependence of
     $\vec{J}_{sys}$ comes from the breaking of translational invariance by
     quenched disorder, and its exact form of $\vec{J}_{sys}$ depends on
     the microscopic dynamics of the model. The {\it diffusive} current,
     ${\vec J}_{diff}=D(\r)\cdot\nabla [\rho(\r,t)-\rho_0(\r)]$, involves a
     quenched diffusion tensor, $D(\r)$. The {\it noise},
     $\vec{\epsilon}(\r,t)$, is to mimic, on a mesoscopic scale, the
     stochastic nature of the evolution.  It is usually assumed to be
     Gaussian distributed and $\delta$-correlated in space and time with
     vanishing spatial and temporal averages.

     To obtain the time evolution of the density fluctuations
     $\rt\equiv\rho(\r,t)-\rho_0(\r)$, we expand $\vec{J}_{sys}$ in powers
     of $\rt$ as
     $\vec{J}_{sys}(\rho_0(\r),\r)+\vec{c}(\r)\rt+\vec{\lambda}(\r)\rt^2\cdots$
     and put it in (\ref{eq:continuity}). This results in
\begin{equation}
\partial_t\rt =
\nabla\cdot[D(\r)\cdot\nabla\rt-\vec{c}(\r)\rt-\vec{\lambda}(\r)\rt^2
\cdots-\vec{\epsilon}(\r,t)]
\label{eq:dynamics-ddim}
\end{equation}

     In one spatial dimension, the above reduces to (with $\r$ replaced
     with $x$) the form
\begin{equation}
\rtt=\partial_x[D(x)\rtx -c(x)\rt-\lambda(x)\rt^2\cdots-\eta(x,t)]
\label{eq:dynamics-1dim}
\end{equation}
     which was considered in \cite{prl97}.  In steady state, the
     time-averaged current must be independent of $x$.  As both $J_{diff}$
     and $\epsilon(x,t)$ vanish under time averaging, the constraint to be
     satisfied is
\begin{equation}
\partial_x\langle J_{sys}(x)\rangle=0
\label{eq:constraint}
\end{equation}
      which is important to account for, as the coefficients in $J_{sys}$
      are explicitly space-dependent, {\it i.e.},
      $\partial_x[\lambda(x)\langle \rt^2 \rangle +...]$ must vanish.

      In their attempt to study a continuum model which describes the
      DASEP, Becker and Janssen (BJ) \cite{Janssen} write the current $J'$
      in powers of $\phi(x,t) \equiv \rho(x,t)-\rho$, the density
      fluctuation away from the {\it overall} particle density $\rho$ in
      the system. In one dimension, the form quoted \cite{DDS} for the
      DASEP is
\begin {equation}
J'=(1-2\rho)\phi(x,t)-\phi^2(x,t)+\eta(x)
\label{eq:BJ}
\end{equation}
      where $\eta(x)$ is an additive quenched noise term. Since the
      time-averaged density $\rho_0(x)$ varies in space, it is evident that
      $\phi$ has a nonzero expectation value $\langle\phi(x,t)\rangle =
      \rho_0(x)-\rho$. As we have seen in Section III, $\rho_0(x)$, and
      thus $\langle \phi(x,t) \rangle$, can show strong variations,
      especially in the density-segregated regime. Spatial constancy of the
      average current demands that the time average of the right hand side
      of (\ref{eq:BJ}) must satisfy
\begin{equation}
\partial_x[(1-2\rho)\langle\phi(x,t)\rangle-\langle\phi^2(x,t)\rangle+\eta(x)]
=0.
\label{eq:BJconstr}
\end{equation}
      Even when this condition is satisfied, it is not entirely clear that
      this continuum theory actually represents the DASEP in one dimension.
      BJ argue that if $\partial J/\partial \rho \ne 0$, then disorder is
      irrelevant on the large scale, a conclusion that is supported by
      \cite{prl97,unpubl}. However, their conclusion, that the condition
      $\partial J/ \partial \rho =0$ holds only if $\rho={1 \over 2}$, does
      not seem to be correct, at least for the DASEP in one dimension; we
      have seen in Section IV, there is an entire range of densities $({1
      \over 2}-\Delta \le \rho \le {1 \over 2}+\Delta)$ where $\partial
      J/\partial \rho$ vanishes, associated with segregation of
      density on a macroscopic scale.

 \section{Equivalent interface models in one dimension}
  \label{sec:height}
      
\begin{figure}[tb]
\begin{center}
\leavevmode
\psfig{figure=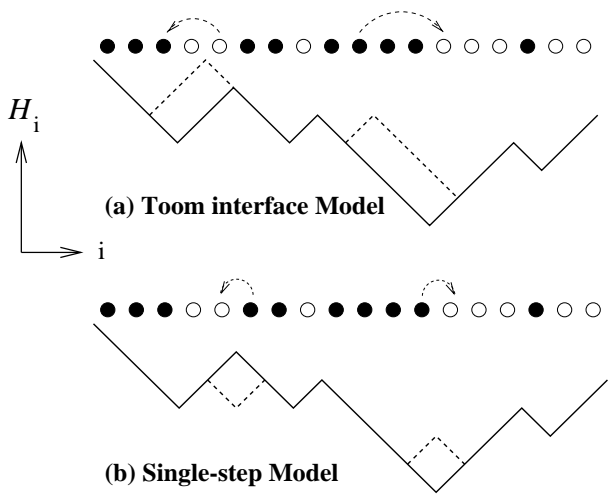,width=8cm}
\end{center}
\narrowtext
\caption{ Mapping between driven particle systems in $d=1$  and 
growing interfaces. (a) Toom Interface dynamics corresponding to the
GDPP, (b) Single-step model corresponding to the DASEP. }
\label{fig:map}
\end{figure}

      In 1-$d$, both the DASEP (with or without backbends) and the GDDP --
      for which the maximum occupancy per site is 1 -- are equivalent to
      stochastic growth models of a 1-$d$ interface moving in a 2-$d$
      medium.  Corresponding to each particle-hole configuration $\{n_i\}$
      is assigned an interface profile $\{H_i\}$ through $H_i=\sum_{j\le i}
      (1-2 n_i)$ \cite{solids}. Pictorially this means that each particle
      corresponds to a $-45^\circ$ downward line segment, while a hole
      corresponds to an upward one (Fig.~\ref{fig:map}). Thus, clusters of
      particles and holes translate to $\pm 45^\circ$ slope segments, and
      the interface has a mean slope which vanishes when the particle
      density is $1/2$.  Away from half-filling, periodic boundary
      condition for the lattice becomes a helical boundary condition for
      the interface.  Junctions between adjacent particle and hole clusters
      correspond to corners in the interface profile.

      Evolution of the interface is dictated by the dynamics of the
      corresponding particle system. The GDDP corresponds to the slice-wise
      motion of segments of a Toom interface in the low noise limit
      \cite{Toom}, while the DASEP corresponds to the corner-flip `single
      step' growth model \cite{solids} (Fig.~\ref{fig:map}).  In both cases
      particle movement to the right (or hole move to left in GDDP)
      corresponds to local forward growth (deposition) of the interface
      while a leftward particle move (e.g. in DASEP with backbends)
      corresponds to local backward growth (evaporation). The quenched jump
      rates for the particle moves implies {\it columnar} disordered
      growth-rates for the interface: the normal growth rate of the
      interface at a fixed $i$ is the same irrespective of the location of
      the interface at successive times.  In the long time limit, the mean
      local forward speed of the interface is the same at all points
      along the interface -- being proportional to the spatially constant
      steady-state current of the corresponding particle model.

\begin{figure}[tb]
\begin{center}
\leavevmode
\psfig{figure=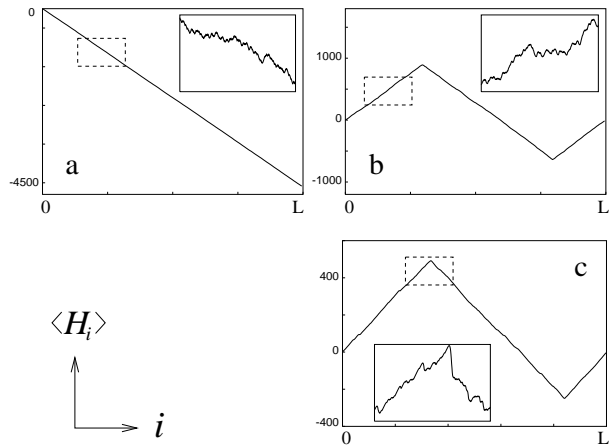,width=8cm}
\end{center}
\narrowtext
\caption{Interface morphology in the three regimes: (a) Uniformly moving
interface with a uniform slope, (b) interface with large
segments of different slopes moving with a non-vanishing speed, and (c)
interface with large segments of different slopes moving with a speed which
vanishes in the thermodynamic limit. The insets show blow-ups of regions
enclosed in the dashed boxes.}
\label{fig:interface}
\end{figure}

      We now discuss the various qualitatively different regimes which
      arise in the interface growth models. Figure \ref{fig:interface}
      shows time averaged steady-state interface profiles $\langle
      H_i\rangle$ corresponding to the three regimes of driven particle
      systems illustrated in Fig.~\ref{fig:regimes1} in Section
      \ref{sec:introduction}.  In all the three cases we started from an
      initially uniform profile -- corresponding to a random
      distribution of particles on the lattice.
      
      Figure \ref{fig:interface}a is an interface with a non-zero mean
      tilt, which has net forward growth rate at all points. The interface
      has a uniform slope on a macroscopic scale and moves with a finite
      non-zero speed preserving its mean tilt. This case corresponds to the
      {\it Homogeneous} regime depicted in Fig.~\ref{fig:regimes1}a. On a
      microscopic scale the interface has frozen-in roughness 
      (Fig.~\ref{fig:interface}a, inset) corresponding to the microscopic
      shocks in the steady-state density profile of
      Fig.~\ref{fig:regimes1}a.

      If the mean tilt vanishes, but the interface still has a net forward
      growth rate at all points, then the initial uniform profile at $t=0$
      coarsens into large segments of different mean slopes at long times
      (Fig.~\ref{fig:interface}b). These segments have frozen roughness on
      microscopic scales, similar to the non-zero tilt case
      (Fig.~\ref{fig:interface}b, inset). The interface moves with a finite
      speed preserving its mean shape and mean vanishing tilt. This
      corresponds to the {\it Segregated-Density } regime of
      Fig.~\ref{fig:regimes1}b.

      In addition to the frozen roughness on the microscopic scales, we can
      define the kinetic roughness as the equal-time mean-square height
      fluctuations around the steady-state profile. We consider the
      zero-mean height variables $h_i(t)\equiv H_i(t)-\langle H_i\rangle$
      defined in (\ref{eq:heightdef}), and define the roughness exponent
      $\alpha$ through ${\overline{\langle(h_{i+r}-h_i)^2\rangle}}\sim
      r^{2\alpha}$.  As discussed in Subsections \ref{sec:DDPP}D,E and
      \ref{sec:DASEP}F, the roughness exponent $\alpha=1/2$, in both the
      cases above.

       In Fig.~\ref{fig:interface}c, which corresponds to the {\it
      Vanishing-Current} regime of Fig.~\ref{fig:regimes1}c, the profile
      resembles that corresponding to the Segregated-Density regime in that
      it has large segments of different mean slopes. However, in this case
      the interface is stationary in the thermodynamic limit -- reflecting
      the vanishing of the steady-state current in the particle system.  In
      interface language, this situation can be visualized as a case of
      interface growth where there are local stretches of the interface
      having a net backward growth (evaporation) rate. Though, on the
      average, there are more forward-growth regions than backward-growth
      ones, in the limit of large system size, arbitrarily long evaporation
      stretches effectively pin the interface.
            
      Turning now to continuum description of the dynamics of these
      interfaces, at least in the cases where the mean speed of growth is
      non-zero, the sum in the definition of the height variable is
      replaced by an integral of the coarse-grained particle density,
      $H(x,t)=\int^x[1-2\rho(x',t)] dx'$. The growth equation for $H(x,t)$
      is governed by (spatially) integrated one dimensional version of the
      continuity equation (\ref{eq:continuity}): $\partial_t
      H(x,t)=J(\partial_x H,x,t)$. The fluctuations $h(x,t)\equiv
      H(x,t)-H_0(x)$in $H$ around the steady-state profile $H_0(x)\equiv
      \langle H(x,t)\rangle$, are governed by
\begin{equation}
\htt= D(x)\htxx - c(x)\htx + {1\over 2}\lambda(x)\htx^2\cdots + 2\eta(x,t)
\label{eq:height}
\end{equation}
obtained by integrating (\ref{eq:dynamics-1dim}). The absence of any additive
quenched spatial noise term in (\ref{eq:height}) is due to the spatial
constancy of the growth speed of the interface dictated by the same
constraint on the steady-state current. 
In this respect (\ref{eq:height}) differs from the model discussed
in \cite{krug} where such a term arises naturally due to absence of any
such constraints. As can be readily verified by power counting, an additive
quenched columnar term is highly relevant in the RG sense and leads to much
rougher interfaces than the $\alpha=1/2$ interfaces described by
(\ref{eq:height}).

\section{Conclusion}
\label{sec:conclusion}

      In this paper, we have studied the stationary current-carrying states
      of driven lattice gas models with quenched disordered hopping rates.
      The principal results are of two types --- first the exact
      determination of the steady states for a class of disordered models,
      and second the demarcation of distinct regimes of behaviour on
      macroscopic length scales, as a result of disorder.  In this section,
      we briefly review these results, and discuss some related open
      problems.
 
      The steady states of a family of disordered models --- the disordered
      drop-push process (DDPP), and related models --- have been found in
      all dimensions by an application of the condition of pairwise
      balance.  The result is a product measure state, with site-dependent
      weights, reflecting the microscopic disorder in the model.  The
      current has been computed as well.  The system is characterized
      by a strictly uniform current density, and a coarse-gained particle
      density that is approximately uniform.  On a macroscopic scale, the
      state is homogeneous.

      Disorder can lead to macroscopically non-homogeneous states, as
      in the 1-$d$ disordered asymmetric simple exclusion process
      (DASEP).  Our numerical and mean-field results show that a
      macroscopically density-segregated state occurs in the DASEP
      model with no backbends, for densities in a finite range around
      half-filling.  The origin of density separation is traceable to
      the existence of a largest current that can be carried by a
      stretch of weak bonds.  This low current can be sustained in the
      rest of the lattice only by separating the density into distinct
      large and small values in macroscopic regions of the lattice.

      Backbends introduce a third type of possible behaviour in one
      dimension.  Like the stretch of weak forward bonds in the
      density-segregated regime, a backbend rate-limits the current,
      leading to density segregation.  However, there is an important
      difference: the longer the stretch of weak forward bonds, the closer
      is the current to a finite asymptotic value; by contrast, the longer
      the stretch of reverse-biased bonds, the smaller the current --- it
      decreases exponentially fast with backbend length.  Since the
      probability of occurrence of a backbend decreases exponentially with
      its length, the result is a current that decreases as a
      bias-dependent power of the overall size of the lattice.

      The crucial physical point which underlies the behaviour in each of
      these regimes is the requirement that the steady-state current be
      constant at all points in the system. Continuum field-theoretical
      approaches must ensure that this local constraint is respected; while
      this is automatically assured for translationally invariant systems,
      it may need special care to guarantee in disordered systems.

      It would be of interest to generalize our results to higher
      dimensions and also to include interactions between particles at
      different sites. A few scattered results along these lines are
      available. \hfill \break \noindent (i) For the drop-push class
      of problems, we have seen in Section \ref{sec:DDPP} that the
      exact steady state even in higher dimensions is characterized by
      inhomogeneous product measure. On large scales, this leads to a
      homogeneous state. \hfill \break \noindent (ii) The transport of
      particles with hard-core interactions through the infinite
      cluster of a randomly diluted lattice above the percolation
      concentration has been studied \cite{RB,BR}.  In a certain
      regime of dilution, backbends act as local traps, but unlike the
      one-dimensional case considered in Section \ref{sec:Sinai},
      there exist infinitely long paths on which the length of every
      backbend is less than a fixed value \cite{BR2,RB,BR}. The
      sub-network of such paths is expected to carry a current which
      then remains finite in the thermodynamic limit. There is thus no
      vanishing-current regime in this system. \hfill \break \noindent
      (iii) With attractive interactions between particles, the driven
      lattice gas system with nearest-neighbour hopping is known to
      undergo phase separation below a certain temperature
      \cite{Katz,DDS}. A numerical simulation showed that the addition
      of a low concentration of blocked sites did not alter the
      critical behaviour of this system \cite{Lauritsen}.

      More systematic studies of higher-dimensional systems are called for.
      In particular, it would be interesting to know whether
      disorder-induced large-scale inhomogeneities, akin to the phase
      separation found in one dimension, persist in higher dimensions as
      well.

      Acknowledgements: We thank R. E. Amritkar, D. Dhar, S.
Krishnamurthy, R. Lahiri, G. I. Menon and N. Trivedi for useful
discussions.

\appendix
\section{Symmetries of the current in the DASEP}
\label{apx:PHsym}
 
      In this appendix we discuss two types of symmetry transformations in
      the DASEP which leave the steady-state current invariant.  (i) The
      first, which involves flipping the directions of all jumps and
      interchanging all particles and holes, can be proved exactly. Under
      these operations, the steady-state weights of particle configurations
      and the magnitude of the current are shown to remain invariant.  (ii)
      The second type of symmetry is restricted to 1-$d$ systems in which
      only one way jumps are allowed on each bond. The symmetry
      transformation consists of flipping the direction of jump on each
      bond.  In this case, we have convincing numerical evidence that the
      current is invariant, though there appears to be no simple
      relationship between steady-state weights of various particle
      configurations. We have no general proof of this result.

\begin{figure}[tb]
\begin{center}
\leavevmode
\psfig{figure=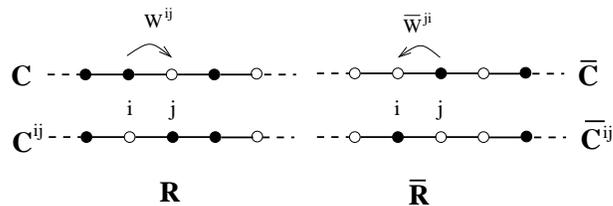,width=8cm}
\end{center}
\narrowtext
\caption{Invariance of the current under $R\rightarrow\Rbar$ and 
$\rho\rightarrow 1-\rho$ for the DASEP in 1-$d$. Jump directions on every
bond are reversed from $R$ to $\Rbar$. Configurations
$\calC,{\overline{\calC}}$ are related by particle-hole interchange.  So
are ${\calC}^{ij}$ and ${\overline{\calC^{ij}}}$.}
\label{fig:PHsym}
\end{figure}

      (i) {\it Easy direction flip and particle-hole interchange}: Consider
      the DASEP, with quenched disordered, unequal forward and backward
      hopping rates on each bond.  Below we explicitly deal with the
      one-dimensional case, but the results are easily generalizable to
      higher dimensions. Let $R$ denote a particular realization of
      disorder and $\Rbar$ the realization obtained from $R$ by flipping
      the easy direction of jumps on all the bonds: If $R$ denotes the set
      of transition rates \{$W^{ij}$\}, then $\Rbar$ corresponds to the set
      \{$\Wbar^{ij}=W^{ji}$\}. Also, let us denote by $\calC$ and ${\Cbar}$
      two particle configurations related to each other by particle-hole
      interchange. Clearly, if $\calC$ is an allowed configuration of the
      system at filling $\rho$, then ${\Cbar}$ corresponds to a filling
      $1-\rho$, and thus there is a one-to-one correspondence between the
      configurations at the two fillings. Now, let $\calC^{ij}$ be the
      configuration obtained from $\calC$ by exchanging the occupation
      numbers at the sites $i$ and $j$.  It is easy to see that the two
      transition rates, $W_\rho(\calC\rightarrow\calC^{ij})$ in realization
      $R$, and $\Wbar_{1-\rho}(\Cbar\rightarrow\Cijbar)$ in realization
      $\Rbar$ are equal (Fig.~\ref{fig:PHsym}), i.e. the transition
      matrices ${\cal W}_\rho$ and $\overline{\cal W}_{1-\rho}$ in the two
      realizations have identical entries.

      Identification of the two ${\cal W}$-matrices implies that the
      invariant measures $P_\rho$ and $\overline{P}_{1-\rho}$ satisfy
\begin{equation}
P_\rho({\calC})=\overline{P}_{1-\rho}({\Cbar}).
\end{equation}
      Using this, together with the identity
      $n_i({\calC})=1-n_i(\Cbar)$, we can relate $n$-point
      correlations at fillings $\rho$ and $1-\rho$. In particular,
      the site densities at the two fillings are related as
\begin{equation}
\langle n_i\rangle_{\rho,R} = 1-\langle n_i\rangle_{1-\rho,\Rbar}.
\end{equation} 
       Further, the steady-state currents in $R$ and $\Rbar$ at the two
       fillings are equal in magnitude:
\begin{eqnarray}
J_0(\rho,R)
&\equiv &W^{ij}\langle n_i(1-n_j)\rangle_{\rho,R}
          -W^{ji}\langle n_j(1-n_i)\rangle_{\rho,R}\nonumber\\ 
&=&{\Wbar^{ji}}\langle n_j(1-n_i)\rangle_{1-\rho,\Rbar} 
           -{\Wbar^{ij}}\langle n_i(1-n_j)\rangle_{1-\rho,\Rbar} \nonumber \\
&=& -J_0(1-\rho,\Rbar).
\label{eq:strong_phsym}
\end{eqnarray}
       The negative sign in the last step reflects the fact that the
       direction of the current is opposite in the two systems.

       In the case of the FSM (subsection \ref{sub:FSM}) the realizations
       $R$ and $\overline{R}$ are identical (apart from a
       reflection). Hence in this case the above arguments imply that the
       magnitudes of the currents at the two fillings $\rho$ and $1-\rho$
       are equal. A similar result holds for the single defect bond case
       studied by Janowsky and Lebowitz \cite{Janowsky}.

       For the DASEP, with or without backbends, (\ref{eq:strong_phsym})
       has the corollary that the disorder averaged currents at the two
       fillings $\rho$ and $1-\rho$ are equal.

      (ii) {\it Easy direction flip only (for 1-$d$, forward hopping)}:
      Above, we found that the current is invariant when the filling is
      changed from $\rho$ to $1-\rho$, provided that the disorder
      realization is changed from $R$ to $\Rbar$. Here we observe (based on
      numerical evidence) that in 1-$d$, with only forward hopping, the
      result is also true for realization $R$ on its own,  i.e.
\begin{equation}
	J_0(\rho,R)=J_0(1-\rho,R). 
\label{eq:phsym2}
\end{equation}
	In view of (\ref{eq:strong_phsym}) this is equivalent to 
\begin{equation}
	J_0(\rho,R)=-J_0(\rho,\Rbar).
\label{eq:phsym3}
\end{equation}
      The claim is easily verified for the single particle (or single hole)
      case using an explicit form for the current:
      $J_0=(\sum_{i}\alpha^{-1}_{i,i+1})^{-1}$ \cite{BD}.  Also,
      (\ref{eq:phsym3}) is true at $\rho=1/2$ since (\ref{eq:phsym2}) is an
      identity at this filling.
  
      We do not have a proof for (\ref{eq:phsym2}) or (\ref{eq:phsym3}) in
      the general case, but they seem to be borne out numerically.  For
      instance, we studied the validity of (\ref{eq:phsym3}) for a system
      of size $L=6$ with $N=2$ particles. We took $R$ to be the set
      $\{W^{i,i+1},i=1,\cdots,6\}=\{\half,\half,1,1,\half,1\}$.  The
      invariance of the current is verified up to $1$ part in $10^8$. We
      also studied the steady-state probabilities of all the $^6C_2$
      particle configurations for each of $R$ and $\Rbar$, both by Monte
      Carlo and by Lanczos iteration of the stochastic evolution
      operator. There seems to be no straightforward correspondence between
      the two sets of steady-state weights. This suggests that there should
      be a proof of the invariance of the current which does not rely on
      identifying the weights of individual configurations.

\end{multicols} 


\begin{thebibliography}{99}
\bibitem{well_distbn}{This set of numbers can in
    general may be {\it any} set of positive numbers. In the disordered
case, they are chosen from a probability distribution. See, for example,
subsection \ref{subsec:2rate} where explicit results are obtained if these
rates are chosen from a binary distribution. }
\bibitem{angle_avg}{Here the angular brackets $\langle\cdot\rangle$ denote 
      time average in the steady state which, in view of the ergodicity of
      the dynamics, is identical to the expectation value with respect to
      the invariant measure.}
\end{thebibliography}
\end{document}